# AMR MAGNETIC SENSORS IN FLUX EXPULSION STUDIES*

A. Netepenko†, Fermi National Accelerator Laboratory, Batavia, USA

*Abstract*

Magnetic flux expulsion properties of the superconducting material such as bulk niobium, widely used for the radio-frequency cavity fabrication, substantially affect the performance characteristics of the cavities. The quality factor of the SRF resonators can be significantly compromised due to the presence of the trapped flux vortices causing additional RF energy losses in the material.

Large number of experiments have been carried out by different research groups to establish the correlation of the flux trapping in niobium cavities with the presence of impurities in niobium as well as various surface treatment methods. Majority of these experiments utilize commercially available cryogenic fluxgate magnetic sensors to measure the field before and after the niobium transition to the superconducting state to quantify the amount of flux trapped.

One disadvantage of the typically used fluxgates is the size of the sensing volume. As an example, the Bartington F and G type cryogenic fluxgates have a sensing core length of about 30mm, which is comparable to the curvature radius of the cavity walls and hence the magnetic field lines curvature radius after the expulsion. Thus, the measured field value needs to be corrected to account for the sensor effective averaging over the sensing volume. In case of the sharper geometries, for instance if the flux expulsion to be measured on the edge of the rectangular niobium flat sheet with a thickness of ~5mm, the use of the fluxgate would be impractical.

Sensors with much smaller core volume need to be utilized for the described experimental setup. One of the candidates for this task is an anisotropic magnetic resistance (AMR) sensor, which has proven to be operational at cryogenic temperatures and has significantly smaller sensing volume, with the sensing element characteristic size of less than 1mm.

A prototype magnetic field measurement system based on the AMR sensors has been designed, which can be utilized for the flux expulsion studies beyond the scope of the SRF cavity geometries. The principles of Wheatstone bridge measurement applied to the anisotropic magnetic resistance sensors are discussed and the realisation of the data acquisition system is described in detail. System capabilities and limitations are investigated. Magnetic field expulsion measurement results for the rectangular flat niobium samples are presented.

## AMR BRIDGE MEASUREMENT

A system of four resistors assembled in a circuit shown in Figure 1. conventionally called a "full bridge" allows to measure the relative resistance change by measuring the voltage drop across the bridge ($\Delta V = V_{out+} - V_{out-}$) while applying a fixed source voltage $V_{cc}$.

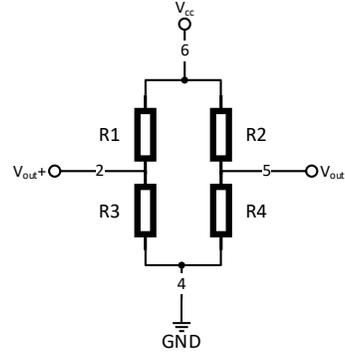

Figure 1. Full bridge (Terminals numbered according to the SENSITEC AFF755B pinout [1]).

In case when all four resistors have equal nominal resistance $R$ with some perturbation $\Delta R_i$, i.e.

$$R_i = R + \Delta R_i, \quad (1)$$

such that $\frac{\Delta R_i}{R} \ll 1$, the voltage across the bridge in first order approximation can be expressed as:

$$\frac{\Delta V}{V_{cc}} = \frac{\Delta R_2 - \Delta R_1 + \Delta R_3 - \Delta R_4}{4R} . \quad (2)$$

Ferromagnetic materials exhibit resistance variation in presence of external magnetic field, the effect called anisotropic magnetoresistance. The effect is a result of *s-d* electron scattering cross section variation depending on the angle between the current and magnetization of the material. Due to the spin-orbit coupling and spin interaction with external magnetic field applied to the sample the electron's *d* orbitals effectively change their orientation resulting in a scattering cross section variation, that in turn leads to the change of the resistance. This effect is utilized in AMR sensors for magnetic field measurement.

The relative variation of the resistance of a ferromagnet satisfies the relation [2]:

$$\frac{\Delta R}{R} \sim \cos^2 \Theta , \quad (3)$$

where Θ is an angle between magnetization and an electric current direction:

$$\cos \Theta = \frac{(\vec{M} + \vec{H}) \cdot \vec{I}}{|\vec{M} + \vec{H}| \cdot |\vec{I}|}, \quad (4)$$

here $\vec{M}$ is an internal magnetization, $\vec{I}$ is a current, and $\vec{H}$ is an external magnetic field being measured.

Resistors in AMR sensor have a "barber pole" structure which leads to the current direction making 45° agnel with the magnetization (Figure 2.) Current flows perpendicular to the highly conductive plates embedded into a ferromagnetic layer.

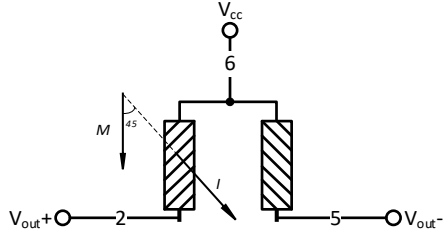

Figure 2. "Barber Pole" resistors in AFF755B.

In the coordinate system where the ferrite magnetization is parallel to $y$ axis, the three vectors from the Equation 4 can be written as:

$$\vec{M} = (0, \pm M); \ \vec{H} = (H_x, H_y); \ \vec{I} = \left(\pm\frac{I}{\sqrt{2}}, \frac{I}{\sqrt{2}}\right) \quad (5)$$

This would lead Equation 2 being transformed into:

$$\frac{\Delta V}{V_{cc}} \sim \frac{H_x(H_y \pm M)}{H_x^2 + (H_y \pm M)^2}. \quad (6)$$

Since the saturation magnetization of the permalloy films typically used in AMR sensors is of the order of $kG$, while the measured magnetic fields are of the order of $G$ or less, meaning $\left(\frac{H_{x,y}}{M}\right)^2 \ll 1$, thus:

$$\frac{\Delta V}{V_{cc}} \sim \frac{H_x}{\pm M}. \quad (7)$$

The voltage measured across the bridge is proportional to the component of the external magnetic field perpendicular to the internal magnetization vector:

$$\frac{\Delta V}{V_{cc}} = \alpha \frac{H_x}{\pm M} + \frac{V_{offset}}{V_{cc}}, \quad (8)$$

where $\alpha$ is a sensitivity of the sensor and additional offset term is originated from the bridge internal imbalance.

To compensate for the offset the magnetization can be flipped up and down periodically and the difference will give a corrected measurement.

## DATA ACQUISITON SYSTEM

For a bridge measurement data acquisition system an existing module solution can be utilized, such as NI-9237 C Series Strain/Bridge Input Module from National Instruments. This module has an internal voltage source ($V_{cc}$) as well as an ADC to measure the voltage across the bridge ($\Delta V$) ang gives the $\frac{\Delta V}{V_{cc}}$ as an output. The module requires a Compact DAQ chassis, NI cDAQ-9174 model was used in our setup. For the control signal pulse generation fed to magnetization flipping coil circuit a digital module NI-9401 was used, placed in the same chassis to allow a synchronized flipping and measurement. Schematics of the data acquisition system can be provided by request.

To understand the measurement capabilities of the developed system one needs to investigate the hardware limitations foremost. Main parameters of AMR sensors from SENSITEC are listed in Table 1.

The sensor noise level for frequencies $< 1Hz$ declared by SENSITEC in [1] is of the order of $0.2\mu V/V$ at room temperature, which corresponds to a $\sim 0.25 mG$ for typical sensitivity (one should note that this noise level is declared for the operation without magnetization flipping, and factor of $\sqrt{f}$ should as well be applied for higher frequency acquisition).

Table 1. AFF755B Specifications

| Symbol | Parameter | Min. | Typ. | Max. | Unit |
|---|---|---|---|---|---|
| $V_{cc}$ | Supply voltage | 1.2 | 5.0 | 9.0 | V |
| $R_B$ | Bridge resistance | 2.2 | 2.5 | 2.8 | $k\Omega$ |
| $S$ | Sensitivity (in range $\pm 2\ G$) | 1.037 | 1.196 | 1.356 | $\frac{\mu V/V}{mG}$ |
| $I_F$ | Flip current (required) | $\pm 150$ | | | mA |
| $R_F$ | Flip coil resistance | | 1.5 | 2 | $\Omega$ |

In typical flux expulsion experiments the magnitude of applied and measured magnetic field is of the order of $100\ mG$, which would give us an AMR response of $\sim 0.120\ mV/V$. This is roughly $2^8$ times smaller than the whole range ($50 mV/V$), hence we effectively using only $16\ bits$ of the module ADC, and amplification of the bridge response would be beneficial but not implemented yet. The resolution is significantly higher than the AMR noise, $50\frac{mV}{V}/2^{24} = 2.98\frac{nV}{V}$, which corresponds to $\sim 3.6\ \mu G$ (while the noise level is of the order of $mG$).

Important parameters of the bridge measurement module NI-9237 are listed in the table below.

Table 2. NI-9237 Module Specifications

| Parameter | Condition | Value | Unit |
|---|---|---|---|
| Sample Rate | | 50 | $\frac{kS/s}{ch}$ |
| Resolution | | 24 | bits |
| Range | | $\pm 25$ | $\frac{mV}{V}$ |
| Input Noise | $V_{cc} = 5V$, $f_{acq} = 100\ Hz$ | 40 | $\frac{nV}{V}$ |

The noise levels are observed to be significantly higher for the magnetization flipping operation, especially at the cryo-temperatures possibly due to the increase of magnetic coercivity of permalloy material used in AMR. This might be restrictive for our use, but the implementation of the flipping method was necessary to observe that limitations, and some technical details of the flipping system follow.

SENSITEC AMR sensors have integrated flipping coil, by sending the current pulses of $\geq 150\ mA$ in forward and reverse direction one can flip the internal magnetization of the sensor. A recommended pulse duration is $\sim 1\mu s$ with $\sim 1ms$ pause between pulses. The current polarity and pulse duration control is realized with the use of L6202 DMOS Full Bridge Driver and the NI-9401 Digital Module. For the proper operation L6202 requires $\sim 36V$ supply voltage, a $100\Omega$ potentiometer used to adjust the current flowing through the flip coils, and the logic signals are supplied to the Enable, Input1 and Input2 pins of the chip to enable the current flow (101 – positive polarity, 110 – negative polarity, 000 – no current). One could run the data acquisition on NI-9237 with maximum rate, perform the flipping by generating proper logic signals on NI-9401 with required duty cycle and do a post processing on the obtained data, subtraction of the plateau value before and after the flip.

Another approach is to synchronize the flipping signals and acquisition such that it is known during the acquisition itself which points belong to the top and which to the bottom plateau. This task appeared to be non-trivial in our hardware configuration due to various limitations of the module's synchronization methods in LabView cDAQ paradigm, but the solution was found.

The common time base is essential for synchronization, and 12.8 $MHz$ NI-9237 module clock is used as a common clock. The module sampling rate is derived from the sampling clock as:

$$f_s = \frac{f_{timebase} \div 256}{n},$$

where $n$ – integer, and $n \in [1, 31]$. Since we need to produce digital pulses of $\sim 1\mu s$ with NI-9401, the time base divisor $\tilde{n}$ equal to 12 can be used in its clock settings (for this module there is no additional divisor of 256). Between flipping up and down we should have $N \sim 1000$ its clock cycles to provide a time gap of $\sim 1ms$ between flips. Assuming we would take $n_m$ measurements between flips one can write:

$$\frac{f_{timebase} \div 256}{n} \cdot \frac{1}{n_m} = \frac{f_{timebase}}{\tilde{n}} \cdot \frac{1}{N}.$$

Suitable solution would give us a combination:
$\tilde{n} = 12;\ n = 12; n_m = 3;\ N = 768;$

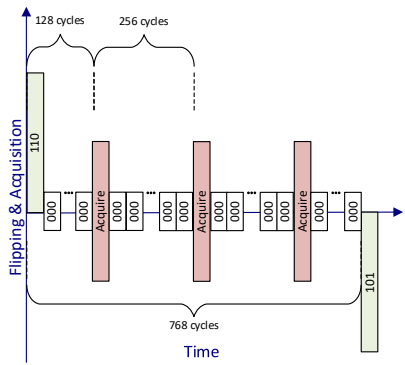

Figure 3. Flipping pulses and acquisition timing.

Writing an array consisting of 1x110, 767x0, 1x101 and 767x0 into the buffer of NI-9401 for continuous generation and triggering the accusation of NI-9237 with a digital output start trigger of the chassis, with a small delay to place three acquisition points evenly on the plateau between flips, we achieve the synchronized acquisition and magnetization flipping.

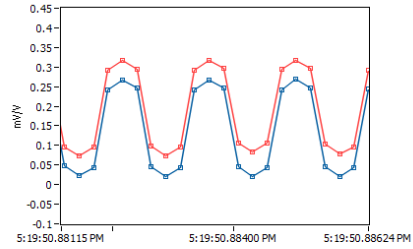

Figure 4. Two AMRs data with synced flipping.

Raw data acquisition rate with the described settings is $\frac{12.8MS/s \div 256}{12} \cong 4.166\ kS/s$, the offset corrected data rate then is 1/6 of this value, or $\sim 0.7\ kS/s$. The corrected data can be averaged on the go and effectively smaller rates can be utilized for lowering the noise level.

## SENSORS CALIBRATION

AMR sensors can be calibrated by placing them in varying magnetic field of Helmholtz coils next to the fluxgate used as a reference. As was mentioned earlier the magnetization flipping technique for the offset compensation does not work at cryogenic temperatures, so both sensitivity and offset of the sensor need to be measured.

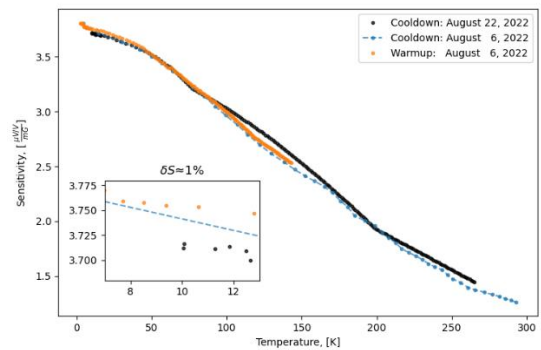

Figure 5. AMR sensitivity as a function of temperature for multiple cooldowns (~280$K$-1.5$K$).

Sensitivity deviation of the order of 1% was observed between several cooldowns from room temperature to ~1.5K. Unfortunately, the offset behavior is much less predictable and prevents us from being able to use the calibration for absolute field value determination, only the field change can be measured with AMRs and described methodic (offset mitigation remains an open question).

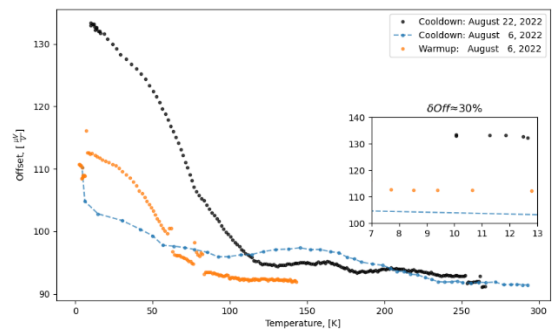

Figure 6. AMR offset as a function of temperature.

## FLAT Nb SAMPLE EXPULSION

Flat niobium sample flux expulsion measurement could allow pre-cavity fabrication evaluation of a material flux trapping properties. Since a sample is placed in a close to uniform magnetic field, in a presence of additional fluxgate the AMR sensors can be used despite the unknown offset.

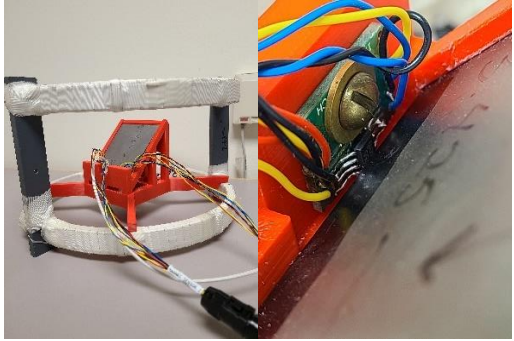

Figure 7. Flat Nb sample setup with two AMRs and one fluxgate at the edge of the sample.

Expulsion simulation carried out with COMSOL package shows substantial advantage of the AMR sensors due to the smaller sensing volume.

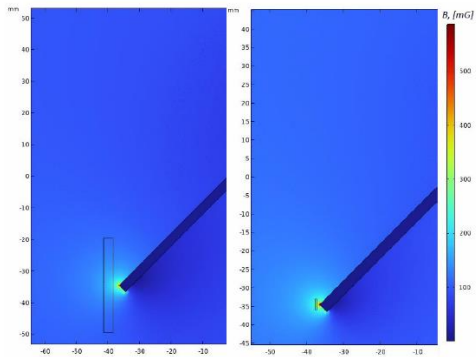

Figure 8. Fluxgate (left) and AMR (right) field distribution from flat sample expulsion.

Post expulsion magnetic field increase seen by the AMR is significantly higher than fluxgate volume averaged field.

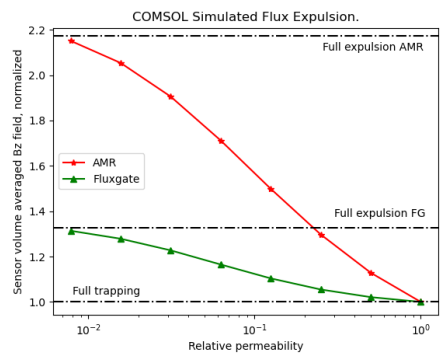

Figure 9. AMR vs Fluxgate measured field enhancement for varying sample expulsion strength.

Experimental results shown on the Figure 10 agree with the simulation-based expectation in terms of the fluxgate to AMR performance comparison, the flux expulsion of the used sample was quite poor (almost full trapping).

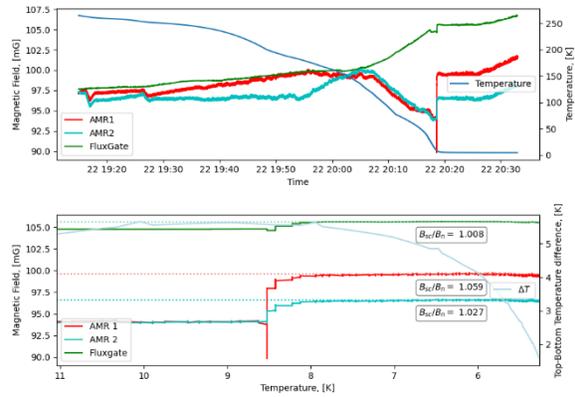

Figure 10. Flat Nb sample expulsion data.

As can be seen from the bottom graph the field value measured by the AMR is different from the fluxgate due to the sensors offset variation between the cooldowns. The field rise during the expulsion is much more pronounced as seen by the AMRs than the fluxgate due to the smaller sensing volume.

## CONCLUSION

AMRs are suitable for the magnetic field change measurement at cryogenic temperature (~1% accuracy achieved). Flat Nb sample expulsion measurement was completed with definitive AMR advantage over the fluxgates.

Small AMR sensing volume can be utilized for trapped flux spacial distribution measurement system design.

Fast AMR response could allow to observe the flux dynamics during the superconducting transition or cavity quenching.

Mitigation of the sensors offset, and accurate measurement of the absolute magnetic field value remain an open question.

## ACKNOWLEDGEMENTS

I would like to thank R. Pilipenko (Fermilab) for his support in DAQ system hardware and software development, and M. Martinello (SLAC) for the flat sample expulsion idea proposal and valuable discussions.